\documentclass[aps,pra,superscriptaddress,twocolumn]{revtex4}
\usepackage{natbib}
\usepackage{epsfig}
\usepackage{bm,epstopdf}
\usepackage{graphicx,times}
\usepackage[usenames]{color}
\usepackage[dvipsnames]{xcolor}
\usepackage{amssymb,amsmath,bbm,amsfonts,amsthm,subfigure}
\newcommand{\bs}[1]{\boldsymbol{#1}}

\newcommand{\ket}[1]{|{#1}\rangle}
\newcommand{\bra}[1]{\langle{#1}|}

\graphicspath{{figure/}}

\begin{document}
\title{Quantum dynamics of a macroscopic magnet operating as environment 
of a mechanical oscillator} 
\author{C.Foti}
\affiliation{Dipartimento di Fisica, Universit\`a di Firenze,
      Via G. Sansone 1, I-50019 Sesto Fiorentino (FI), Italy}
\affiliation{INFN Sezione di Firenze, via G.Sansone 1,
      I-50019 Sesto Fiorentino (FI), Italy}
\author{A. Cuccoli}
\affiliation{Dipartimento di Fisica, Universit\`a di Firenze,
       Via G. Sansone 1, I-50019 Sesto Fiorentino (FI), Italy}
\affiliation{INFN Sezione di Firenze, via G.Sansone 1,
     I-50019 Sesto Fiorentino (FI), Italy}
\author{P. Verrucchi}
\affiliation{Istituto dei Sistemi Complessi,
       Consiglio Nazionale delle Ricerche,
       via Madonna del Piano 10,
       I-50019 Sesto Fiorentino (FI), Italy}
\affiliation{Dipartimento di Fisica, Universit\`a di Firenze,
       Via G. Sansone 1, I-50019 Sesto Fiorentino (FI), Italy}
\affiliation{INFN Sezione di Firenze, via G.Sansone 1,
       I-50019 Sesto Fiorentino (FI), Italy}

\begin{abstract}
We study the dynamics of a bipartite quantum system in a way such 
that its formal description keeps holding even if one of its parts 
becomes macroscopic: the problem is related with the analysis of the 
quantum-to-classical crossover, but our approach implies that the whole 
system stays genuinely quantum. Aim of the work is to understand {\it 
1)} if, {\it 
2)} to what extent, and possibly {\it 3)} how, the evolution of a macroscopic 
environment testifies to the coupling with its microscopic quantum 
companion. To this purpose we consider a magnetic environment made of 
a large number of spin-$\frac{1}{2}$ particles, coupled with a quantum 
mechanical oscillator, possibly in the presence of an external 
magnetic field.  We take the value of the total 
environmental-spin $S$ constant and large, which allows 
us to consider the environment as one single macroscopic system, and 
further deal with the hurdles of the spin-algebra via approximations 
that are valid in the large-$S$ limit. 
We find an insightful expression for the propagator of the whole system, 
where we identify an effective "back-action" term, i.e. an operator 
acting on the magnetic environment only, and yet missing in the absence of the 
quantum principal system. This operator emerges as a time-dependent 
magnetic anisotropy whose character, whether uniaxial or planar, also 
depends on the detuning between the level-splitting in the spectrum of 
the free magnetic system, induced by the possible presence of the 
external field, and the frequency of the oscillator. The time-dependence 
of the anisotropy is analysed, and its effects on the dynamics of 
the magnet, as well as its relation with the entangling evolution of the 
overall system, are discussed.
\end{abstract}

\date{\today}

\maketitle

\section*{Introduction} \label{s.Intro} 

\noindent 

%rilevanza dello studio dell'interazione fra un sistema genuinamente 
%quantistico ed un altro macroscopico:

For almost the whole last century the problem of how a principal quantum 
system ($\Gamma$) behaves when interacting with a macroscopic 
environment ($\Xi$) has been considered assuming the latter to be a 
classical system. If this is the case, a quantum analysis of 
how the two subsystems evolve due to 
their reciprocal interaction is hindered, which is quite a severe 
limitation since macroscopic environments are the tools by which we 
ultimately extract information about, or exercise control upon, any 
microscopic quantum system 
\citep{Schlosshauer07, Rivas2012, BreuerP02, PaladinoEtal02}. 
In particular, the effects of the presence 
of $\Gamma$ on the way $\Xi$ evolves (often referred to as "back-action" 
in the literature) have no place in the description, and entanglement 
between the twos is neglected. 

Recently, however, hybrid schemes in which micro- and macroscopic 
systems coexist in a quantum device have been considered in different 
frameworks, from the analysis of foundational issues via optomechanical 
setups, to quantum thermodynamics or nanoelectronics 
\citep{LoFrancoEtal13, XuEtal13, Matsumoto15, KippenbergV08, VerhagenDWSK12}. 
In fact, it is not completely clear why one should renounce a quantum 
description of a macroscopic system: after all, this 
is nothing but a system made of many quantum particles that, for one 
reason or another, can be described regardless of its internal structure 
as if it were a single object with its own, effective, Hilbert space. 
The exemplary case of such situation is when $\Xi$ is made by a large 
number $N$ of spin-$\frac{1}{2}$ particles and is such that its total 
spin $S$ is a conserved quantity: no matter how large $N$ is, the 
corresponding magnetic environment behaves, in general, as a quantum 
system: this is clearly seen if its total spin equals, say, $S=1/2$ or 
$S=1$. On the other hand, for $S\propto N\to\infty$ a classical-like 
dynamics is expected \cite{Lieb73}, while large-$S$ approximations are 
ideal tools for studying macroscopic, and yet quantum, magnetic systems. 
In general, models that are hybrid in the sense explained above must be 
studied with the toolkit of open quantum systems enriched by specific
accessories for dealing with the macroscopicity of some of their elements.

With this in mind, we here consider a magnetic environment $\Xi$, made 
by a large number $N$ of spin-$\frac{1}{2}$ 
particles, featuring a global symmetry that guarantees the 
total spin $S$ to be a constant of motion. As far as 
$S$ is finite, such magnet is the prototype of a system that exhibits a 
distinct quantum behaviour despite being macroscopic ($N\gg 1$).
The microscopic companion of the magnet is assumed to be a quantum mechanical 
oscillator $\Gamma$, with which $\Xi$ exchanges energy according to a 
model-Hamiltonian that goes beyond the pure-dephasing 
interaction \citep{PalmaSE96, CucchiettiEtal10}.

We address the time evolution of the composite system $\Gamma+\Xi$ by a 
large-$S$ approximation
that represents the macroscopicity of $\Xi$, since $N\ge 2S$ holds,
without totally wiping out its quantum character, since $S$ is finite.
Moreover, such an approximation allows us to deal with the 
complications due to the involved algebra of the spins; in fact, making 
use of recent results \cite{CasasMN12} on the factorization of operatorial 
exponentials, and the Zassenhaus formula\cite{Zassenhaus39,CasasMN12}, 
we obtain a factorized expression for the propagator of the composite 
system and find that, due to the coupling between $\Gamma$ and $\Xi$, a 
specific term appears, effectively representing the back-action of the 
principal system on its environment. 
Indeed, the factorization of the propagator allows us to 
define a free effective Hamiltonian 
which includes the back-action term in the form of 
a time-dependent magnetic anisotropy, whose intensity and character 
(axial or planar) vary, to represent the non-entangling 
component of the dynamics due to the interaction with the underlying 
quantum oscillator.

The work is structured as follows: in Sec.~\ref{s.environment} we define 
the magnetic environment $\Xi$ and briefly discuss the relation between
the large-$S$ condition and macroscopicity.
The principal system $\Gamma$ enters the scene in Sec.~\ref{s.partner}, 
where the Hamiltonian, containing an interaction of Tavis-Cummings 
form \cite{TavisC68}, is introduced. The propagator is evaluated 
in Sec.\ref{s.dancing1}, making use of the Zassenhaus expression in 
the large-$S$ approximation. Results are presented in 
Secs.~\ref{s.effective}-\ref{s.dancing2}, 
and conclusions drawn in Sec.~\ref{s.results}.

\section{The magnetic environment}
\label{s.environment}

Let us consider a magnetic system $\Xi$ made of $N$ spin-$\frac{1}{2}$ 
particles, each described by its Pauli matrices 
$(\hat\sigma_i^x,\hat\sigma_i^y,\hat\sigma_i^z)
\equiv\hat{\bs\sigma}_i$. As we will always understand $\hslash$ 
finite, we can hereafter set $\hslash = 1$.
Be $\hat{\boldsymbol S}\equiv\frac{1}{2}\sum_i^N\hat{\boldsymbol 
\sigma}_i$
the total spin of $\Xi$ and 
$|\hat{\bs{S}}|^{2}\equiv(\hat S ^x)^{2}+(\hat S ^y)^{2}+ 
(\hat S ^z)^{2}=S(S+1)$, with $S$ ranging from $0$ to $N/2$ if $N$ 
is even (from $1/2$ to $N/2$ if $N$ is odd). When $|\hat{\bs{S}}|^{2}$ 
commutes with the propagator, the value $S$ stays constant and $\Xi$ can be 
seen as one single physical system described by the spin operators 
closed under the $su(2)$ commutation relations $[\hat S ^{\alpha},\hat S 
^{\beta}] = i \varepsilon_{\alpha \beta \gamma} \hat S ^{\gamma}$, with 
$\alpha(\beta,\gamma)=x,y,z$.
Notice that taking $|\hat{\bs{S}}|^{2}$ conserved implies 
assuming that a global symmetry exists in the Hamiltonian acting on 
$\Xi$, where "global" means that its generators, amongst which 
$|\hat{\bs{S}}|^{2}$ itself, 
have the same, non-trivial, action on the Hilbert space of any of the 
$\Xi$-components.
One such symmetry characterizes, for 
instance, a system made by $N$ spin-$\frac{1}{2}$ particles, possibly 
distributed on the sites of a ring (see Fig.~1), which are either 
independent or coupled amongst themselves via a homogeneous, isotropic 
(or Ising) nearest-neighbour interaction, $j \sum_i\hat{\boldsymbol 
\sigma}_i\cdot\hat{\boldsymbol\sigma}_{i+1}$ (or
$j \sum_i\hat\sigma^\alpha_i\hat\sigma^\alpha_{i+1}$).

\begin{figure} [h] \centering
\subfigure[]{ \label{f.spinring}
\includegraphics[scale=0.6]{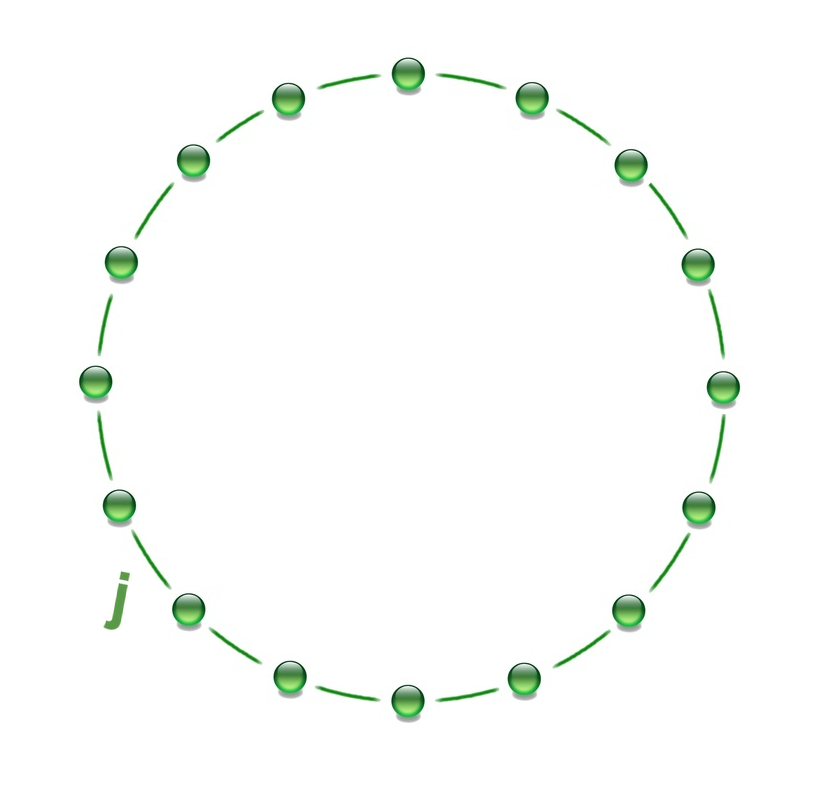}}
\subfigure[]{ \label{f.quantum_partner}
\includegraphics[scale=0.6]{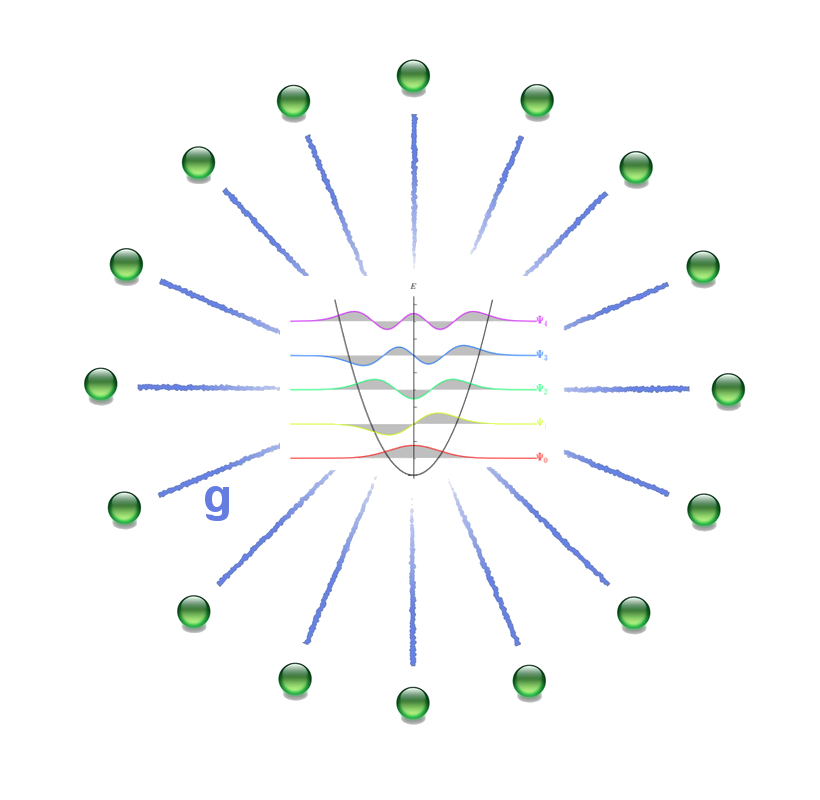} }
\caption{Graphical representation of a magnetic system made of 
distinguishable particles, with equal spin, 
distributed on a ring-shaped 
lattice (referred to as a "spin-ring" in the text). In panel (a) the 
system is isolated and its components interact with each other; in panel 
(b) the system is coupled with a quantum mechanical 
oscillator and its components are independent from each other.}
\end{figure}

Once the total spin is guaranteed a constant value $S$, one can consider 
that $S\to\infty$ is a necessary 
condition for spin systems to behave classically. In fact, without 
entering into the detailed formalism that allows one to consistently 
describe the quantum-to-classical crossover of a
magnetic system~\cite{Lieb73, Yaffe82}, this can be naively understood 
by the following argument: 
defining the normalized spin operator 
$\hat{\bs{s}}{\equiv}\hat{\bs{S}}/S$, it is
%\begin{equation}\label{e.largeS}
$[\hat{\bs{s}} ^\alpha, \hat{\bs{s}}^\beta] = 
i\varepsilon_{\alpha \beta \gamma}
\hat{\bs{s}}^\gamma/S$,
\noindent which implies that $\hat{\bs{s}}$ becomes a classical 
vector in the $S\rightarrow \infty$ limit.
In Sec.~\ref{ss.large-S} we will show how to introduce a large-$S$ 
approximation, essentially based on the above argument.

\section{The quantum partner}
\label{s.partner}
The "spin-ring" introduced in the previous Section, see Fig.~1(a), 
is now identified as the magnetic environment $\Xi$ of a quantum 
mechanical oscillator $\Gamma$, see Fig.~1(b). We choose
the Hamiltonian of the overall system of the form
\begin{equation}
\label{e.star-hamiltonian}
\hat{H}= \omega\hat{a}^{\dagger} \hat{a} + \frac{h}{2} 
\sum_i^N\hat{\sigma}_i^z + 
\frac{1}{2}\sum_i^N g_i(\hat{a} 
\hat{\sigma}_i^{+} + \hat{a}^{\dagger} \hat{\sigma}_i^{-})~,
\end{equation}
\noindent 
where 
%$\omega$ is the free frequency of $\Gamma$, 
$h$ is an 
external field defining the $z$ axis, and 
$\hat\sigma^{\pm}_i\equiv\hat\sigma_i^x\pm i\hat\sigma_i^y$; the 
different $g_i$ are the couplings between each spin of the ring and the 
oscillator. Being $\hbar=1$, for the bosonic 
operators describing the principal system it holds
$[\hat{a},\hat{a}^{\dagger}]=\hat{\mathbb{I}}$.

In order for the model \eqref{e.star-hamiltonian} to describe a 
system whose environment can be made macroscopic, one needs
guaranteeing the existence of a global symmetry such that the total spin 
is conserved. This can be 
accomplished implementing different conditions, amongst which we choose 
$g_i=g~\forall i$, leading to the Tavis-Cummings (TC) model
\citep{TavisC68, Garraway11, BennettEtal13}
\begin{equation}
\hat{H}=
g(\hat{a} \hat S ^+ + \hat{a}^{\dagger} \hat S ^-)
+(\omega \hat{a}^{\dagger} \hat{a} + h \hat S_z)=
\hat Y +\hat X~,
\label{e.TC-hamiltonian}
\end{equation}
where we have defined the free, 
$\hat X\equiv \omega \hat{a}^{\dagger} \hat{a} + h \hat S_z$, and
interacting, 
$\hat Y\equiv g(\hat{a} \hat S ^+ + \hat{a}^{\dagger} \hat S ^-)$, 
terms.
This is an  exactly solvable model~\cite{TavisC68}, and
analytic expressions for its eigenvectors and eigenvalues exist;
however, these expressions are useless if one aims at writing
the propagator in a form that lend for the recognition of
different components in the overall dynamics, which is indeed our goal.
In fact, the TC model is usually  studied taking the bosonic mode 
as the environment, for a principal system which is, in a way or 
another, described by the spin operators $\hat{\bs{S}}$ 
\citep{HaerkonenPM09, FengEtal15}.
If one tries to analyze the TC dynamics 
regarding the spin as the environment, formal problems due to the 
spin-operator algebra for large $S$ emerge, which is the reason 
why this choice most often 
trails behind itself that of a completely classical treatment of the 
environment, resulting in the 
replacement of the Hamiltonian's 
spin operators with a classical field $\bs B(t)$, with "ad hoc" 
time-dependences 
\cite{PaladinoEtal02, PaladinoEtal14, BenedettiEtal13, WoldEtal12}. 
To this respect, we notice 
that describing a quantum system via a time-dependent Hamiltonian 
implies assuming that an 
environment exists, which is not however sensitive to the presence of 
the principal system itself. In fact, the time dependence of the 
field $\bs B(t)$ is arbitrarily chosen and does not change with the 
principal system's evolution, a condition that defines the so called "no 
back-action" approximation. 
On the other hand, if one aims at studying quite the back-action that 
the environment experiences because of its interaction with the 
principal system, it is necessary to consider the TC model with
the spin system described as a genuinely quantum, magnetic environment.

\section{The propagator}
\label{s.dancing1}

The evolution induced by the TC Hamiltonian is 
severely convoluted: not only the free ($\hat X$) and 
interacting ($\hat Y$) terms of Eq.~\eqref{e.TC-hamiltonian} 
do not commute, but the spin-commutation relations further prevent one 
from obtaining usable expressions via the Backer-Campbell-Haussdorff 
formula. In fact, it is quite clear that, as far as the coupling $g$ in 
Eq.\eqref{e.TC-hamiltonian} is finite, any attempt of disentangling
the propagator $\exp(- i \hat{H} t)$ by taking out factors separately 
acting on $\Gamma$ and $\Xi$ will face the 
problem of dealing with infinitely nested commutators.

We take on the problem of studying the evolution
\begin{equation}
\ket{\Psi(t)}=e^{-i\hat H t}\ket{\Psi(0)}=e^{\lambda(\hat Y+\hat 
X)}\ket{\Psi(0)}~,
\label{e.prop}
\end{equation}
%\begin{equation}
%\ket{\Psi(t)}=\e^{-i\hat H t}\ket{\Psi(0)}
%\label{e.propagator}
%\end{equation}
%and 
with $\lambda\equiv -it$,
by means of the left-oriented version of the Zassenhaus formula, so as 
to make the free term $\hat X$ act directly on the 
initial state $\ket{\Psi(0)}$, as will be done in Sec.~\ref{s.dancing2}. 
The left oriented Zassenhaus formula can be written \cite{CasasMN12},
as follows 
\begin{equation} e^{\lambda (\hat Y + \hat{X})} = \cdots 
e^{\lambda^{n} \widetilde{C}_{n}} \cdots e^{\lambda^{3} 
\widetilde{C}_{3}} e^{\lambda^{2} \widetilde{C}_{2}} e^{\lambda \hat Y} 
e^{\lambda \hat{X}} \; ,\label{e.Zassenhaus} \end{equation} 
\noindent 
where $\widetilde{C}_{n} = (-1)^{n+1} C_{n}$ with $n\geq2$, and the 
Zassenhaus operators $C_{n}$ are given in terms of 
\begin{eqnarray} \label{e.commutators}
ad_{\hat{X}}^{0} \hat Y=\hat Y \;\;\;&,&\;\;\;ad_{\hat{X}} \hat 
Y=[\hat{X},\hat Y] \nonumber \\ ad_{\hat{X}}^{k} \hat 
Y&=&[\underbrace{\hat{X},[\hat{X} \;...\;[\hat{X}}_{k-times},\hat 
Y]...]]~, 
\end{eqnarray} 
\noindent 
and the same for $\hat{X} \leftrightarrow \hat Y$. 
In particular it is 
\begin{eqnarray} 
\label{e.Zassenhaus-coefficients} C_{n+1}=\frac{1}{n+1} 
\sum_{i_{0},\;i_{1},\;...\;,\;i_{n}} && 
\frac{(-1)^{i_{0}+i_{1}+\cdots+i_{n}}}{i_{0}! i_{1}! \cdots i_{n}!} 
\nonumber\\ &\cdot & ad^{i_{n}}_{C_{n}} \cdots ad^{i_{2}}_{C_{2}} 
ad^{i_{1}}_{\hat Y} ad^{i_{0}}_{\hat{X}}\hat Y\;,\label{e.Cn}
\end{eqnarray} 
where each $(n+1)$-tuple of non negative integers 
$(i_{0},\;i_{1},\;...\;,\;i_{n})$ must satisfy 
\begin{eqnarray}
\label{e.ntuples} 
&& i_{0}+i_{1}+2i_{2}+...+ni_{n}=n \nonumber\\
&{\rm and}&\label{e.nuple}\\
&& i_{0}+i_{1}+2i_{2}+...+ji_{j}\geq j+1
\;\mbox{for}\;j=0,...,n-1.\nonumber
\end{eqnarray} 
We underline that, as demonstrated in Ref.~\cite{CasasMN12}, 
the 
commutators defining the separate terms of the sum in 
Eq.\eqref{e.Zassenhaus-coefficients} are 
all linearly independent: this means that, once the commutator defined 
by a certain $(n+1)$-tuple has been determined, it is guaranteed that no 
other $(n+1)$-tuple will give the same operator. Moreover, we notice that 
the time-dependence of each exponential in Eq.~\eqref{e.Zassenhaus} 
follows the ordering of the Zassenhaus terms in powers of $t$, so that 
$t^m$ exclusively multiplies $\widetilde C_m$, for all $m$.
As for the order in $g$, it is easily seen that each commutator in 
Eq.~\eqref{e.Cn} is proportional to $g^l$, where $l$ is the 
number of operators $\hat Y$ entering its definition. These features
allow us to monitor the validity of the approximation scheme hereafter 
adopted, as extensively discussed at the end of 
Sec.~\ref{ss.propagator}.\

\subsection{Large-$S$ approximation}
\label{ss.large-S}
In the Introduction we have underlined that one of the features 
that characterizes a system as "environment" is that of being 
macroscopic. We have then seen, in Sec.~\ref{s.environment},
that when dealing with an environment described by spin operators, one 
can consistently implement macroscopicity by choosing a large value of $S$.
On the other hand, if we take a large $S$ and still want to 
mantain the original picture of a quantum system $\Gamma$ interacting with its 
equally quantum environment $\Xi$, we
must require that the interaction Hamiltonian stay finite for
$S\gg 1$, implying that the coupling $g$ in 
Eq.~(\ref{e.TC-hamiltonian}) scales as $1/S$
\cite{notasulcampo}.
Therefore, we take $gS$ constant (in fact we set $gS=1$ in what follows)
and assume
\begin{equation} \label{Approximation}
g^{m} \prod_{i=1} ^{n<m} \hat S ^{\alpha_i}\sim 0 \; ;
\end{equation}
\noindent the symbol "$\sim$" will be hereafter used to explicitely 
remind that condition \eqref{Approximation} is assumed.
It is important to notice that this large-$S$ approximation is utterly 
different from those required for making spin-boson transformations
tractable by truncating square roots of operators, as done when using 
the Holstein-Primakoff or Villain transformations \citep{MattisbookI}. 
In these cases the spin-sphere, i.e. the isomorphic manifold 
of the $su(2)$ algebra, is projected onto a plane or a cylinder, 
respectively, which is parametrized by the usual conjugate coordinates: 
this implies 
that the algebra of the analyzed quantum system is substantially 
altered. On the contrary, Eq.\eqref{Approximation} keeps the 
spin-character of the magnetic operators without modifying 
their associated geometry, so that terms like {\it axial}, {\it 
planar}, {\it pole}, {\it equator}... simultaneously mantain their 
meaning.

Let us now get back to Eq.\eqref{e.Zassenhaus}: in order to obtain the 
operators $\widetilde C_{n}$, we define
\begin{equation}
\delta\equiv(h-\omega)~~~{\rm and}~~~
\hat{\overline{Y}}\equiv g(\hat{a} \hat S ^{+}-\hat{a}^{\dagger}
\hat S ^{-})~,\label{e.delta-Ybarra}
\end{equation}
use
\begin{eqnarray}\label{e.comXY}
&&[\hat{X},\hat Y]=\delta\hat{\overline{Y}}~~~,~~~
[\hat{X}, \hat{\overline{Y}}]=\delta \hat Y \\
&&[\hat Y, \hat{\overline{Y}}]= -2g^{2} (2 
\hat{\textit{a}}^{\dagger} \hat{\textit{a}} \hat S _{z} + 
\hat S ^+ \hat S ^-)\;,\nonumber
\end{eqnarray}
and find that, due to condition \eqref{Approximation}, 
only two types of commutators survive:
\begin{eqnarray}
 [\underbrace{\hat{X},[\hat{X}
\;...\;[\hat{X}}_{n-times},\hat Y]...]]&\equiv&
ad^{\;n}_{\hat{X}}\hat Y=\nonumber\\ 
&=&
g\delta^{n}\left(\hat a\hat S^+ + (-1)^{n}
\hat a^\dagger \hat S^-\right)~~~\label{e.surviveone}
\end{eqnarray}
and 
\begin{eqnarray}
{[\hat Y,\underbrace{[\hat{X},[\hat{X} %
\;...\;[\hat{X}}_{(n-1) - times\;, \; n\;even}}\!&{,\!}&\hat Y]...]]]
\equiv ad_{\hat Y} ad_{\hat{X}}^{\;n-1} \;\hat Y=
\nonumber\\ &=& -2g^{2}\delta^{n-1}\hat S ^{+}\hat S ^{-}~.
\label{e.survivetwo}
\end{eqnarray}
This implies, referring to conditions \eqref{e.nuple}, that only the 
following
$(n+1)$-tuples remain in the sum entering 
Eq.\eqref{e.Zassenhaus-coefficients}:
\begin{eqnarray} 
&&i_{0} = n\;\;\mbox{with}\;\;i_{k}=0\;\;\forall k\neq 0~,{\rm and}
\nonumber\\ &&
i_{0}=n-1\;\;,\;\;i_{1}=1\;\;\mbox{with}\;\;i_{k}=0\;\;\forall k\neq
0,1\;. \end{eqnarray}
\noindent Therefore, 
defining $\hat Y^+\equiv g\hat{a}\hat S ^{+}$ and $\hat Y^-
\equiv g\hat{a}^{\dagger} \hat S ^{-}$,
the Zassenhaus operators are found to be:
\begin{eqnarray} \label{e.Zassenhaus-coeff_largeS}
\widetilde{C}_{2m+1}&=& C_{2m+1} \sim\frac{1}{(2m + 1)!}
\delta^{2m} (\hat Y^+ + \hat Y^- ) \nonumber\\ &+&
\frac{2m}{(2m+1)!}  \delta^{2m-1}( -2 g^{2}
\hat S ^{+}\hat S ^{-} )\\
 \widetilde{C}_{2m}&=& - C_{2m} \sim\frac{1}{(2m)!} \delta^{2m-1}
(\hat Y^+ - \hat Y^- )\;.\nonumber
\end{eqnarray}
We underline that the Zassenhaus operators $\widetilde C_2$ and 
$\widetilde C_3$ only contain commutators of the form 
\eqref{e.surviveone}-\eqref{e.survivetwo}, 
meaning that expressions \eqref{e.Zassenhaus-coeff_largeS} 
are exact for $m=1$.
Finally, based on condition \eqref{Approximation}, we will hereafter use
\begin{eqnarray}
[\hat Y^+,\hat Y^-]&=&[g\hat a \hat S^+,g\hat a^\dagger \hat S^-]\sim\nonumber\\
&\sim& g^2\hat S^+\hat S^-\sim g^2\hat S^-\hat 
S^+\sim\label{e.S+S-commute}\\
&\sim& g^2\left[S(S+1) - \hat S _z ^{2}\right]~,\label{e.anisotropy}
\end{eqnarray}
\begin{equation}
[[\hat Y^+,\hat Y^-],\hat Y^\pm]\sim
[[\hat Y^+,\hat Y^-],h\hat S^z]\sim 0~,
\label{e.cnumber}
\end{equation}
and hence, as far as the evaluation of the propagator \eqref{e.prop}
is concerned,
\begin{equation}
e^{\hat Y^++\hat Y^-}\sim
e^{\hat Y^+}
e^{\hat Y^-} e^{-\frac{1}{2}g^2\hat S^+\hat S^-}~.
\label{e.BHC}
\end{equation}
We underline that $[\hat Y^+,\hat Y^-]$ does not vanish, despite 
condition \eqref{Approximation} being enforced, because of the 
non-commutativity of $\hat a$ and $\hat a^\dagger$, an evidence that we 
will comment further at the end of Sec.\ref{ss.back-action}.

\subsection{Propagator}
\label{ss.propagator}
We now get back to Eq.~\eqref{e.prop} and 

\noindent{\it i)} Isolate $e^{\lambda\hat X}$:
\begin{eqnarray}
&&\exp\left[\lambda(\hat Y + \hat{X})\right]\sim\nonumber\\
&&\cdots
\exp\left[\frac{\lambda^{n} (\delta)^{n}}{ n!\delta} ( \hat Y^+
-(-1)^{n} \hat Y^- )\right]
\cdots\nonumber\\
&&\times\exp\left[\lambda(\hat Y^+ + \hat Y^- )\right]\nonumber\\
&&\times\exp\left(-2g^2 K_{1\delta}(\lambda)
\hat S ^{+}\hat S ^{-}\right)
\exp\left(\lambda \hat{X}\right)~;\label{e.i}
\end{eqnarray}
{\it ii)} Factorize the exponentials containing both 
$\hat Y^+$ and $\hat Y ^-$:
\begin{eqnarray} \label{e.ii} 
&&\exp\left[\lambda(\hat Y + \hat{X})\right]\sim\nonumber\\
&&\cdots\exp\left(\frac{\lambda^{n}\delta^{n}}{n!\delta}\hat Y^+\right)
\exp\left[-\frac{(-\lambda)^n\delta ^n}{ n!\delta} \hat Y^-\right]
\nonumber\\
&&\cdots\exp\left(\lambda\hat Y^+\right)
\exp\left[-(-\lambda) \hat Y^-\right]\nonumber\\
&&\times\exp\left(-2g^2 K_{2\delta}(\lambda)
\hat S ^{+}\hat S ^{-}\right)
\exp\left(\lambda \hat{X}\right)~;
\end{eqnarray}
{\it iii)} Group together the $\hat Y^-$ ($\hat Y^+$):
\begin{eqnarray}
&&\exp\left[\lambda
(\hat Y + \hat{X})\right]\sim\nonumber\\
&&\exp\left[\frac{1}{\delta} \sum_{n\geq 1} \frac{\lambda^{n}
\delta ^{n}}{n!} \hat Y^+\right] \exp\left[- \frac{1}{\delta}\;
\sum_{n\geq 1} \frac{(-\lambda)^{n} \delta ^{n}}{n!}
\hat Y^-\right] \nonumber \\ 
&&\times\exp\left(K_{3\delta}(\lambda) g^2
\hat S ^{+}\hat S ^{-}\right)
\exp\left(\lambda \hat{X}\right)~.
\label{e.iii}\end{eqnarray} 
The second to last exponential in Eqs.~(\ref{e.ii}), and 
(\ref{e.iii}), accounts for the commutators introduced via 
Eq.\eqref{e.BHC} while first 
factoring, and then swapping, all the exponentials of $\hat Y^+$ 
and/or $\hat Y^-$;
the explicit forms of the functions $K_{*\delta}(\lambda)$, as well as 
the details of the above three steps, are given in 
Appendix.

We are now in the position of summing up the series in Eq.~(\ref{e.iii}), 
which are equal to $(e^{\pm\lambda\delta}-1)$, and finally get the 
global propagator in the form
\begin{eqnarray}
\exp(-i\hat H t)&\sim&\label{e.propfin}\\
&\exp&\left\{g\left[f_\delta(t) \hat{a} \hat S ^{+} -f_\delta^{*}(t) 
\hat{a}^{\dagger}
\hat S ^{-}\right]\right\}\label{e.interaction}\\
\times&\exp&\left[g^{2}G_\delta(t)\hat S ^{+}\hat S ^{-}\right]
\label{e.back-action}\\
\times&\exp&\left(-it \hat{X}\right)~,\label{e.free}
\end{eqnarray}
where the real time $t=i\lambda$ is back, 
$f_\delta(t)\equiv(e^{-it\delta}-1)/\delta$,
and the function $G_\delta(t)\equiv K_3(-it)-|f_\delta(t)|^2/2$
is pure imaginary (as shown in Appendix).

The conditions under which the above form of the propagator holds
are determined as follows. Since products of $n$ spin operators have been 
neglected if multiplied by $g^m$ with $m>n$, according to
condition (\ref{Approximation}), it must be $g\ll 1$, consistently with 
the large-$S$ assumption with $gS$ finite.
As for the time-dependence, we remind that the condition 
\eqref{Approximation} does not affect
$\widetilde C_2$ and $\widetilde C_3$, and Eq.\eqref{e.Zassenhaus} 
with Zassenhaus coefficients from Eqs.\eqref{e.Zassenhaus-coeff_largeS} 
is exact up to the third order in $t$. Moreover, 
we notice that terms linear in whatever spin-operator $\hat S^*$ appear,
through steps {\it i)}-{\it iii)}, as $g^nt^n\hat S^*$ and  
are only kept for $n=1$, which is a valid choice 
if $gt\ll 1$ i.e, as we have set $gS=1$, $t\ll S$.
On the whole, the condition $t\ll S$, with $S$ large, 
defines the proper time-scale in which our results hold true.

\subsection{Back-Action} 
\label{ss.back-action}

The most relevant feature of the above expression 
(\ref{e.propfin}-\ref{e.free}) is 
the appearance of the term $g^2 G_\delta(t)\hat S ^{+}\hat S ^{-}$ 
that has no equivalent in the original 
Hamiltonian and, despite regarding the magnetic system only, is 
effectively generated (as made evident by its being proportional to the 
square of the coupling) by its interaction with the mechanical 
oscillator, thus standing as the type of back-action we were actually 
aiming at describing. In fact, if one reviews the way the above term is 
obtained, it becomes clear that condition \eqref{Approximation} can 
be enforced without wiping the back-action off the global 
dynamics, if and only if $[\hat a,\hat a^\dagger]$ does not vanish
(see comment at the end of Sec.\ref{ss.large-S}). In other terms, it 
is the quantum character of the oscillator that keeps the back-action 
alive in the large-$S$ limit, i.e. when the magnet becomes macroscopic.

In order to better understand the effects of the $\hat S^+\hat S^-$ 
term, we remind that $G_\delta(t) \in \Im$, notice that 
Eq.\eqref{e.cnumber} ensures that $[g^2G_\delta(t)\hat S^+\hat 
S^-{-}it\hat X]$ commutes with itself at different times,
and set
\begin{equation}\label{e.At}
g^2G_\delta(t)=-i\int_0^t A_\delta(\tau)\,d\tau~,
\end{equation}
with $A_\delta(t)$ real: this allows us to define
the effective time-dependent free Hamiltonian 
\begin{equation}
\hat X_\delta^{\rm eff}(t)\equiv A_\delta(t)\hat S^+ \hat S^- +\hat X~,
\label{e.Xeff}
\end{equation}
such that
\begin{eqnarray}
\exp(-i\hat H t)&\sim&\label{e.effprop}\\
&\exp&
\left\{g\left[f_\delta(t) \hat{a} \hat S ^{+} -f_\delta^{*}(t) 
\hat{a}^{\dagger} \hat S ^{-}\right]\right\}
\label{e.effinteraction}\\
\times&\exp&\left[-i\int_0^t \hat X_\delta^{\rm eff}(\tau)\,d\tau\right]~.
\label{e.efffree}
\end{eqnarray}
As for the interaction term, we notice that despite being
$
f_\delta(t)=-i\int_0 ^{t} d\tau\; e^{-i\delta \tau}
$
one is unable to find an effective time-dependent interaction Hamiltonian, 
$\hat Y^{\rm eff}_\delta(t)$ analogous to $\hat X^{\rm eff}_\delta(t)$,
as the argument of the exponential \eqref{e.effinteraction} 
does not commute with itself at different times, unless $\delta=0$. 
If this is the case, however, $f_0(t)=-it$ and the 
exponential (\ref{e.effinteraction}) transforms into the 
propagator of $g(\hat a\hat S^+ + \hat a^\dagger\hat S^-)$; moreover, 
from the general form of $G_\delta(t)$ given in Appendix, one
easily finds $G_0(t)=0$, implying that a genuine interaction picture for 
$\Psi$ emerges; in other terms, when the free evolutions of $\Gamma$ 
and $\Xi$ are resonant there is no back-action whatsoever, and
information is not transferred from one system to the other.

The emergence of an effective Hamiltonian for the magnetic system 
containing a term $\propto \hat S^+\hat S^-$ is consistent with the 
results of Ref.\citep{BennettEtal13}, where however different 
approximations are considered that do not include any 
time dependence for such effective term.

\section{Effective environmental hamiltonian}
\label{s.effective}
The operator $\hat X^{\rm eff}_\delta(t)$ can be  
interpreted as the sum 
of the original free Hamiltonian for the bosonic mode, 
$\hat H_\Gamma=\omega \hat a^\dagger \hat a$, plus an effective, 
time dependent, environmental one
\begin{eqnarray}
\hat H^{\rm eff}_\Xi(t)&\equiv&h\hat S^z+A_\delta(t)\hat S^+\hat S^-\nonumber\\
&\sim&h\hat S^z-A_\delta(t)(\hat S^z)^2-\epsilon_\delta(t)~,
\label{e.Heff_xi}
\end{eqnarray}
where we have used Eqs.~(\ref{e.S+S-commute}-\ref{e.anisotropy}) and 
set $\epsilon_\delta(t)=A_\delta(t)S(S+1)$.
In this way, we see that the presence of 
$\Gamma$ makes the environment feel an effective magnetic anisotropy 
$-A_\delta(t)(\hat S^z)^2$ that favours or hinders the alignment of 
its spin along the quantization axis, depending on the sign of 
$A_\delta(t)$. The time-dependence of $A_\delta(t)$ 
represents the continuous updating of the back-action, which is 
ruled by the energy exchange between $\Gamma$ and $\Xi$.
In particular, it is $A_\delta(t)\propto t^2+{\mathcal{O}}(t^4)$
(from the analytical expression of $G_\delta (t)$ in Appendix and
Eq.~\eqref{e.At}), meaning that there 
exists an initial time-interval during which the environment is not affected by 
the presence of $\Gamma$ in any way other than that due to their 
explicit interaction. 

After some time, however, the energy exchange 
implied by that very same interaction becomes so costly to cause 
a reaction that switches on the back-action, in the form of a 
magnetic anisotropy.  We analyze this fenomenology 
in some details with the help of Figs. 2-4, where lines fade if the 
conditions that guarantee the validity of our results ($t\ll 
S$) are not rigorously met.

\begin{figure} [h] \centering
\includegraphics[scale=0.55]{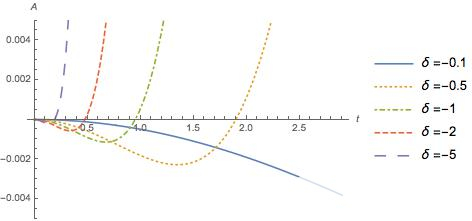}
\caption{Effective anisotropy $A_\delta(t)$ as a function of $t$, for 
$S=10$ and different values of negative $\delta$, as indicated.  
The curve for $\delta=-0.1$ fades when the 
validity of the results is not fully under control (specifically for $t>S/4$).}
\label{f.At_S10+delta}
\end{figure}

In Fig.~\ref{f.At_S10+delta} we show the time evolution of the effective 
anisotropy $A_\delta(t)$ for $S=10$ and some
negative values of $\delta$: We see that 
$A_\delta(t)$ initially works against the magnetic field, favoring 
the spread of the environmental magnetic moment on the $xy$-plane.
As time goes by, however, $A_\delta(t)$ changes its sign (for $t\simeq 
1/|\delta|$), thus preventing the dynamics to freeze by reverting its 
character into an easy-axis one.
%The rapid increase of $A_\delta(t)$ which is observed at larger 
%times is not in the range of validity of our results and we believe it 
%has no physical meaning. 
As for the dependence on the detuning, 
we observe that $A_\delta(t)$ stays negative for 
longer time and 
displays a deeper minimum for smaller values of $|\delta|$:
we understand this evidence by noticing that small values of the 
detuning entail energy scales for the two subsystems comparable to each 
other, which implies that the environment closely follows the beat of 
its quantum partner for a longer time-interval. 

\begin{figure} [h] \centering
\includegraphics[scale=0.55]{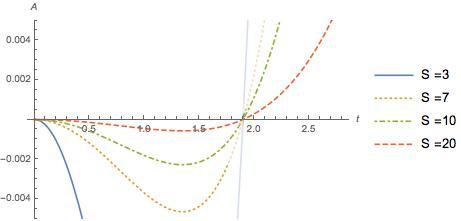}
\caption{Effective anisotropy $A_\delta(t)$ as a funcion of $t$, for 
$\delta=-0.5$ and different values of $S$, as indicated.  
Lines as in Fig.\ref{f.At_S10+delta}.}
\label{f.At_delta05+S}
\end{figure}

In Fig.~\ref{f.At_delta05+S} we set $\delta=-0.5$ and consider 
different values of $S$: we find that $|A_\delta(t)|$ decreases as $S$ 
increases, to represent the growing inefficacy of $\Gamma$ in altering 
the dynamics of its environment as this becomes macroscopic.
In fact, as briefly discussed in the Introduction, a classical-like 
dynamics, with no back-action at all, must characterize
the magnetic environment when $S\to\infty$, which conforms to the 
vanishing of the anisotropy observed for large $S$ in the plot.

\begin{figure} [h] \centering
\includegraphics[scale=0.55]{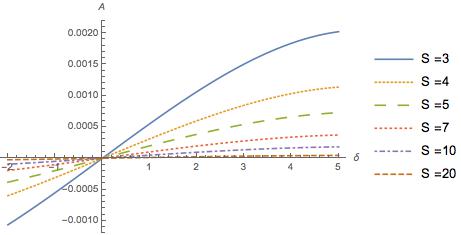}
\caption{Effective anisotropy $A_\delta(t)$ as a function of 
$\delta$, for $t=0.1$ and different values of $S$, as indicated.}
\label{f.Adelta_t01}
\end{figure}

In the above comments, and figures 2-3, we have considered the case of 
negative detuning, $h<\omega$.
The opposite case, $h>\omega$, trivially follows from
$A_\delta(t)=-A_{-\delta}(t)$, as seen from the expression of 
$G_\delta(t)$ in the Appendix, as well as from Fig.~4, where we see that 
the effective anisotropy at a given time is an odd function of 
$\delta$, for all values of $S$.

\section{The evolved state}
\label{s.dancing2}
The factorized form of the propagator 
Eqs.~(\ref{e.propfin}-\ref{e.free})
allows us to identify, amongst the overall effects of the 
interaction between $\Gamma$ and $\Xi$, those that do not 
generate entanglement between the twos.
This is better seen and understood considering the 
evolved state for the entire system $\Psi = \Gamma \cup \Xi$, 
assuming its initial state $\ket{\Psi(0)}$ be separable, i.e.
$\ket{\Psi(0)}=\ket{\Gamma}\otimes\ket{\Xi}$ (we will hereafter 
understand the symbol $\otimes$ whenever possible).
From Eqs.~(\ref{e.effprop}-\ref{e.efffree}) we get
\begin{eqnarray}\label{e.evo_state}
 \ket{\Psi(t)}&=& e^{-i \hat{H}t} \ket{\Gamma}\ket{\Xi}\sim
\nonumber\\
&\sim &e^{g(f_\delta(t) \hat{a} \hat S ^{+} -f_{\delta} ^{*}(t)\hat{a}^{\dagger} 
\hat S ^{-})}
e^{-i\omega\hat{a}^\dagger \hat{a} t}\ket{\Gamma}\nonumber\\
&&\otimes\;e^{-i\int_0^t \hat X^{\rm eff}_\Xi(\tau)\,d\tau}\ket{\Xi}\sim \nonumber\\
&\sim &e^{g(f_\delta(t)\hat{a} \hat S ^{+} -f_{\delta} ^{*}(t) \hat{a}^{\dagger} 
\hat S ^{-})}
\ket{\Gamma(t)}\ket{\tilde{\Xi}(t)}~,
\end{eqnarray}
\noindent where $\ket{\Gamma(t)}=e^{-i\omega\hat{a}^\dagger \hat{a} 
t}\ket{\Gamma}$
and $\ket{\tilde{\Xi}(t)}=\exp[-i\int_0^t \hat X^{\rm 
eff}_\Xi(\tau)\,d\tau]\ket{\Xi}$
describe the free evolution of the bosonic system and the 
effective free evolution of the magnetic one, respectively. 
We have used the notation $\ket{\tilde{\Xi}(t)}$
to underline that while $\ket{\Gamma(t)}$ does not depend on the 
interaction between $\Gamma$ and $\Xi$, the evolution of 
$\ket{\tilde{\Xi}(t)}$ is induced not only by the free Hamiltonian 
$h\hat S_z$, but also by the back-action $g^2 G_\delta (t) \hat 
S^+\hat S^-$ that follows from its coupling with $\Gamma$.

In the above expression \eqref{e.evo_state} we can recognize
a sort of interaction picture with two distinct
rotating frames, 
one for the principal system and one for the environment, that do not 
move independently. In particular, it is the latter that 
changes its pace according to the continuous update of the 
non-commuting components of the environmental 
magnetic moment implied by an interaction of the TC form.
It is worth noticing, to this respect, that the spin commutation 
relations, that in our case are the obstacle to the adoption of an exact 
interaction picture and the reason why an approximation scheme must be 
adopted, effectively manifest themselves in the non trivial 
time-dependence of the back-action, to represent their essential role in
the quantum dynamics generated by the Hamiltonian \eqref{e.TC-hamiltonian}.

\begin{figure} [h] \centering
\subfigure{ 
\hspace{-6mm}
\includegraphics[scale=0.35]{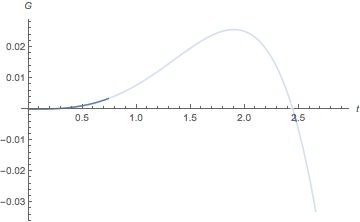}}
\vspace{-5mm}
\includegraphics[scale=0.55]{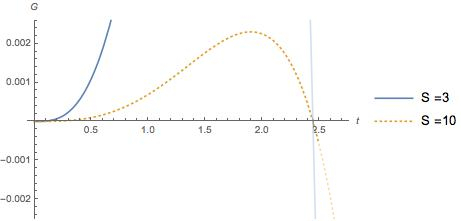}
\caption{ The back-action $g^2|G_\delta(t)|$ for $\delta=-0.5$ 
and different values of $S$; the inset shows the $S=3$ case in its 
proper plot-range. Lines as in 
Fig.~\ref{f.At_S10+delta}.}
\label{f.Gdelta}
\end{figure}

Reminding that $G$ is pure imaginary, in Fig.~\ref{f.Gdelta} we plot  
$g^2|G_\delta(t)|$ as a function of time for $\delta=-0.5$ and $S=3,10$.
Its behaviour qualitatively shows that the back-action has its maximum 
effect, at least as far as the time-interval where our approximation 
holds, for $t\simeq 1/|\delta|$ and vanishes 
for $t>\simeq 1/|\delta|$, no matter the value of the $S$.

Let us finally focus our attention upon the environmental 
reduced density matrix; 
writing the projector $\varrho(t)=\ket{\Psi(t)}\bra{\Psi(t)}$ and 
tracing out 
the $\Gamma$-degrees of freedom, we get
\begin{equation}
\varrho_\Xi(t) \sim \sum_\gamma 
\hat O^{{\gamma,\Gamma}} _{{\Xi}}(t) \ket{\tilde{\Xi}(t)} 
\bra{\tilde{\Xi}(t)}\hat O^{{\gamma,\Gamma}\; \dagger} _{{\Xi}}(t)\;,
\end{equation}
where we have defined the operators 
\begin{equation}\label{e.Kraus}
\hat O^\gamma _\Xi(t)\equiv\hat O^\gamma _\Xi(t;\ket{\Gamma(t)})
\equiv\bra{\gamma}
e^{g(f_\delta(t)\hat{a} \hat S ^{+} -f_{\delta} ^{*}(t) \hat{a}^{\dagger} \hat S ^{-})}
\ket{\Gamma(t)}\;
\end{equation}
and $\lbrace \ket{\gamma}\rbrace$ is an orthonormal 
basis on $\mathcal{H}_\Gamma$.
The set of operators $\lbrace \hat O^\gamma_\Xi(t)\rbrace$ 
acting on the Hilbert space of the environment can be interpreted as one
set of Kraus operators \citep{Nielsen2004}, since the completeness relation
\begin{equation}
\sum_\gamma \hat O^{\gamma\dagger} _{{\Xi}}(t)
\hat O^\gamma_{{\Xi}}(t) = \hat{\mathbb{I}}_\Xi 
\end{equation}
holds for all $t$, as one can easily verify.
% via
%\begin{eqnarray}
%&&\sum_\gamma \bra{\Gamma(t)}
%e^{g(f^*_\delta(t)\hat a^\dagger \hat S ^- -f_\delta (t) \hat a \hat S ^+)}
%\ket{\gamma} \cdot\nonumber\\
%&&\quad\cdot\; \bra{\gamma}
%e^{g(f_\delta(t)\hat{a} \hat S ^{+} -f_{\delta} ^{*}(t) \hat{a}^{\dagger} \hat S ^{-})}
%\ket{\Gamma(t)} =\nonumber\\
%&=&\bra{\Gamma(t)}e^{g(f^*_\delta(t)\hat a^\dagger \hat S ^- -f_\delta (t) \hat a 
%\hat S ^+)} \cdot\nonumber\\
%&&\cdot\; e^{g(f_\delta(t)\hat{a} \hat S ^{+} -f_{\delta} ^{*}(t) \hat{a}^{\dagger}
% \hat S ^{-})} \ket{\Gamma(t)}=\nonumber\\ 
%&=& \hat{\mathbb{I}}_\Xi\;,
%\end{eqnarray}
%Notice that, as the symbol "$\sim$" reminds us, the results
%above holds in the limit of our approximation \eqref{Approximation}.
The fact that the emerging Kraus operators do not depend on 
$G_\delta(t)$ is fully consistent with the fact that the back-action 
does not generate entanglement, as commented above, and rather 
dinamically renormalizes the environmental free Hamiltonian 
$\hat H^{\rm eff}_\Xi(t)$.
We do also notice that, in order for the back-action to have a non 
trivial effect on the environment, the initial state $\ket{\Xi(0)}$ must 
be different from whatever eigenstate of $\hat S^z$, to avoid  
the anisotropy term in $\hat X^{\rm eff}_\delta(t)$ to
affect $\ket{\tilde \Xi(t)}$ only by a phase factor.

\section{Conclusions}
\label{s.results}
In this concluding section we gather the information obtained so far
in order to devise a strategy that make the 
dynamics of $\Xi$ the most sensitive possible to its 
interaction with $\Gamma$. In fact, as mentioned in the Introduction, if 
$\Xi$ is the measuring instrument used for getting information on, or 
exert our control upon, the quantum system $\Gamma$, one such strategy 
might reveal details, or allow a steering precision, otherwise
inaccessible. To this respect, the lesson learnt in this work goes as 
follows. 

1) Detuning: $\delta=h-\omega$ must be finite if one wants to observe 
footprints of $\Gamma$ into an effectively-free evolution of $\Xi$, i.e. 
without further interacting with $\Gamma$ itself. Off-resonance is a 
necessary condition for the back-action to switch on.  

2) Timing: depending on the value of $\delta$ and $S$, there exist a 
finite time interval, that can be well within the range of validity 
of our results as shown in Figs.~\ref{f.At_S10+delta}-\ref{f.Gdelta},
where the back-action is larger, meaning that 
effects of $\Gamma$ on the dynamics of $\Xi$ could 
be more pronounced.

3) Magnetic properties: although our results are obtained in the 
large-$S$ approximation, it is important that $S$ stays finite, to avoid 
the disentangled dynamics of $\Xi$ to be just a silent Larmor 
precession. For the same reason, it is important that $\Xi$ be prepared 
in an initial state which is not an eigenstate of $\hat S^z$: 
significantly, in Ref.~\cite{Foti_Master2015} we have seen that spin 
coherent states \citep{Lieb73, Gilmore72} might be a particularly 
significant choice.

We conclude by mentioning that the method here used for 
implementing the large-$S$ approximation, i.e. making explicit the 
dependence of the spin-algebra on the quanticity parameter 
$1/S$ and then requiring the interaction Hamiltonian to stay finite as such 
parameter drops, is general and might turn useful in studying 
other quantum systems with several interacting components, amongst which a 
macroscopic one, furthermore preserving the geometry of the spin-sphere. 

\begin{acknowledgements}
This work is done in the framework of the Convenzione operativa between 
the Institute for Complex Systems of the Italian National Research 
Council (CNR), and the Physics and Astronomy Department of the 
University of Florence. Financial support from CNR, under the 
Short-Term-Mobility program, is gratefully aknowledged by PV. 
\end{acknowledgements}

\appendix*
\section{}
\label{s.aI}
The results of points {\it i)}-{\it ii)} of Sec.\ref{ss.propagator} are 
obtained by the repeated use of Eq.~(\ref{e.BHC}), realizing in 
Eq.~\eqref{e.i} with
\begin{eqnarray}\label{e.K1_A}
K_{1\delta}(\lambda)&=&\frac{1}{\delta^2}\sum_{m \geq 1} \frac{2m}{(2m+1)!} \lambda^{2m+1} \delta ^{2m+1} \nonumber\\
&=&\frac{1}{\delta^2}\left( -it\delta \cos t\delta + i \sin t\delta 
\right)~,
\end{eqnarray}
and Eq.~\eqref{e.ii} with 
\begin{equation}\label{e.K2_A}
K_{2\delta}(\lambda)=K_{1\delta}(\lambda) - \frac{1}{4\delta^2} \sum_{n\geq 1}(-1)^{n} \left[\frac{\lambda^{n} \delta ^{n}}{ n!}\right]^{2}\;.
\end{equation}

As for the point {\it iii)}, in order to group together all the terms 
proportional to $\hat Y^+$ ($\hat Y^-$), we perform the 
necessary $\hat Y^+\leftrightarrow \hat Y^-$ permutations 
in the infinite product of exponentials entering Eq.~\eqref{e.ii}, and 
get
\begin{eqnarray} \label{e.iii_A}
&&\exp\left[\lambda(\hat Y + \hat X)\right]\sim\nonumber\\
&\sim & \exp\left[\frac{1}{\delta} \sum_{n\geq 1} \frac{\lambda^{n} \delta ^n}{n!} 
\hat Y^+\right] \exp\left[- \frac{1}{\delta}\; \sum_{n\geq 1} \frac{(-\lambda)^n \delta ^n}{n!} 
\hat Y^-\right] \nonumber \\
&&\times \exp\left[-2g^2 K_{2\delta}(\lambda) \hat S^+ \hat S^-\right]  \nonumber \\
&&\times \exp\left[\zeta_\delta(\lambda) g^2 \hat S^+ \hat S^-\right] 
\exp\left(\lambda \hat{X}\right)~,
\end{eqnarray}
where $\zeta_\delta(\lambda)$ is the coefficient resulting from the commutators 
$[\hat Y^+,\hat Y^-]$, introduced while moving all the $\hat Y^-$ to the right.

In order to determine $\zeta_\delta(\lambda)$, we consider 
\begin{eqnarray} \label{e.BCH_A}
&&e^{\mu \hat Y^-} e^{\pi \hat Y^+} \sim e^{\mu \hat Y^- + \pi \hat Y^+ + \frac{1}{2} \mu \pi [\hat Y^-, \hat Y^+]} = \nonumber\\
&&=e^{\pi \hat Y^+} e^{\mu \hat Y^-}  e^{- \mu \pi [\hat Y^+,\hat Y^-]} = e^{\pi \hat Y^+} e^{\mu \hat Y^-}  e^{- \mu \pi g^{2} \hat S^+ \hat S^-}\;
\end{eqnarray}
and define
\begin{equation}
\pi_n = \frac{\lambda^n \delta ^n}{ n!\delta} \;\;\;\mbox{and}\;\;\;\mu_n= -\;\frac{(-\lambda)^n \delta^n}{ n!\delta}\;,
\end{equation}
\noindent so that the expression from which we will get $\zeta_\delta(\lambda)$ 
(see Eq.~\eqref{e.iii_A}) reads 
\begin{equation}
\cdots \; e^{\mu_{\ell + 1} \hat Y^-}\;e^{\pi_{\ell} \hat Y^+}\;\cdots \; e^{\mu_{3}\hat Y^-}\; e^{\pi_{2} \hat Y^+}\; e^{\mu_{2} \hat Y^-}\;e^{\pi_{1}\hat Y^+} e^{\mu_{1} \hat Y^-} \;.
\end{equation}

We then need to exchange every $\pi_n \hat Y^+$ with all the $\mu_\ell \hat Y^-$ of the following orders, i.e. such that $n>\ell$: after the first permutation we get
\begin{eqnarray}
&&\cdots \; e^{\mu_{\ell + 1} \hat Y^-}\;e^{\pi_\ell \hat Y^+}\;\cdots \; e^{\pi_3 \hat Y^+}e^{\mu_3\hat Y^-} \nonumber\\
&&\underbrace{e^{\pi_2\hat Y^+} e^{\pi_1\hat Y^+}}_{e^{(\pi_1+\pi_2) \hat Y^+}}  
e^{\mu_2\hat Y^-}e^{-\mu_2 \pi_1 g^2 \hat S^+\hat S^-} e^{\mu_1 \hat Y^-}\;,
\end{eqnarray} 
\noindent and one can easily check that successive permutations give the terms
\begin{eqnarray}
& e^{-\mu_2 \pi_1 g^2 \hat S^+ \hat S^-}& \nonumber \\
& e^{-\mu_3 (\pi_1+\pi_2) g^2 \hat S^+ \hat S^-} &\nonumber \\
&  \vdots & \nonumber \\
&  e^{-\mu_{\ell +1} (\pi_1+\pi_2+ \pi_3+...+\pi_\ell)g^2 \hat S^+\hat S^-} \;,&\nonumber
\end{eqnarray}
so that
\begin{eqnarray}\label{e.k_A}
&&\exp\left[\zeta_\delta(\lambda) g^2 \hat S^+ \hat S^-\right] \sim \nonumber\\
&&\sim \exp\left[-\left(\sum_{\ell\geq 2} \mu_\ell 
\sum_{1\leq j < \ell} \pi_j\right) g^2 \hat S^+ \hat S^-)\right] =\nonumber\\
&&= \exp\left[\frac{1}{\delta^2}\left(\sum_{\ell \geq 2} 
\frac{(-\lambda)^\ell \delta ^\ell}{ \ell !}\sum_{1\leq j < \ell}  
\frac{\lambda ^j \delta ^j}{ j!}\right)g^2 \hat S^+ \hat S^-\right]\;.\nonumber\\
\end{eqnarray}

Therefore, from Eq.~\eqref{e.iii}, it is
\begin{eqnarray}\label{e.K3_A}
&&K_{3\delta}(\lambda)=-2K_{2\delta}(\lambda) + \zeta_\delta(\lambda)=\nonumber\\
&=&-2K_{2\delta}(\lambda)+\frac{1}{\delta^2}\sum_{\ell \geq 2} 
\frac{(-\lambda)^\ell \delta ^\ell}{ \ell !}
\sum_{1\leq j < \ell}  \frac{\lambda ^j \delta ^j}{ j!}=\nonumber\\
&=&-2K_{1\delta}(\lambda) + \frac{1}{2\delta^2}
\sum_{n\geq 1}\frac{(-\lambda\delta)^n}{n!}\frac{(\lambda\delta)^n}{n!}+\nonumber\\
&&+ \frac{1}{\delta^2}\sum_{\ell \geq 2} \frac{(-\lambda \delta) ^\ell}{ \ell !}\sum_{1\leq j < \ell}  
\frac{(\lambda \delta) ^j}{ j!}~.
\end{eqnarray}

Being $\lambda=-it$, we notice that $-\lambda = \lambda^*$ and set 
$x=\lambda\delta$, $x^*=\lambda^*\delta$; the last two terms can be written as
\begin{eqnarray}
&&\frac{1}{\delta^2}\left(\frac{1}{2} \sum_{n \geq 1} \frac{x^{*n}}{n!} \frac{x^n}{n!} 
+  \sum_{\ell \geq 2} \frac{x^{*\ell}}{ \ell !}
 \sum_{1\leq j < \ell}  \frac{x ^j}{ j!} \right)=\nonumber\\
&=&\frac{1}{2\delta^2} \sum_{n \geq 1} \frac{x^{n}}{n!} 
\sum_{\ell \geq 1}\frac{x^{*\ell}}{\ell !} -\frac{1}{2\delta^2}
\sum_{n \geq 1} \frac{x^{n}}{n!} 
\sum_{\ell \neq n}\frac{x^{*\ell}}{\ell !}+\nonumber\\
&&+  \frac{1}{\delta^2}\sum_{\ell \geq 2} \frac{x^{*\ell}}{ \ell !}
 \sum_{1\leq j < \ell}  \frac{x ^j}{ j!} \nonumber\\
&=&\frac{1}{2}|f_\delta(\lambda)|^2 + M_\delta(\lambda)~, 
\end{eqnarray}
where, going back to $\lambda$ and $\delta$, the first serie can 
be written in terms of $f_\delta(\lambda)\equiv(e^{\lambda\delta}-1)/\delta$, 
i.e. the function defined in Eq.~\eqref{e.interaction}, and
\begin{eqnarray}
M_\delta(\lambda)&=& -\frac{1}{2\delta^2}\sum_{n \geq 1} \frac{(\lambda\delta)^n}{n!}
\sum_{\ell \neq n}\frac{(\lambda^*\delta)^\ell}{\ell !}+\nonumber\\
&&+ \frac{1}{\delta^2}\sum_{\ell \geq 2} \frac{(\lambda^*\delta)^\ell}{ \ell !}
 \sum_{1\leq j < \ell}  \frac{(\lambda\delta) ^j}{ j!}~.
\end{eqnarray}

The propagator \eqref{e.iii_A} is then
\begin{eqnarray}
&&\exp\left[\lambda
(\hat Y + \hat X)\right]\sim\nonumber\\
&\sim& \exp\left[ f_\delta(\lambda)\hat Y^+\right] 
\exp\left[- f^*_\delta(\lambda)\hat Y^-\right] \nonumber\\ 
&&\times\exp\left(K_{3\delta}(\lambda) g^2
\hat S ^+\hat S ^-\right)
\exp\left(\lambda \hat X\right)\nonumber\\
&\sim& \exp\left[ f_\delta(\lambda)\hat Y^+ 
- f^*_\delta(\lambda)\hat Y^-\right] 
\exp\left(-\frac{1}{2}| f_\delta(\lambda)|^2 g^2 \hat S^+ \hat S^-\right)\nonumber\\ 
&&\times\exp\left(K_{3\delta}(\lambda) g^2
\hat S ^+ \hat S ^-\right) \exp\left(\lambda \hat X\right)~, 
\end{eqnarray}
where in the last step we have used $e^{f_\delta(\lambda)\hat Y^+} 
e^{- f^*_\delta(\lambda)\hat Y^-} \sim e^{f_\delta(\lambda)\hat Y^+ 
- f^*_\delta(\lambda)\hat Y^- - \frac{1}{2} \left| f_\delta(\lambda) \right|^2
[\hat Y^+, \hat Y^-]}$ 
and Eq.~\eqref{e.S+S-commute}. Looking at Eq.~\eqref{e.back-action}, 
we therefore have
\begin{equation}\label{e.G_A}
G_\delta(\lambda) = K_{3\delta}(\lambda)-\frac{1}{2}| f_\delta(\lambda)|^2
= -2K_{1\delta}(\lambda) + M_\delta(\lambda)~.
\end{equation}

We now want to show that the above expression is a pure imaginary one; 
since $K_{1\delta}(\lambda) \in\Im$, this means actually to show that 
$M_\delta(\lambda) \in \Im$. Restoring $x=\lambda\delta$ and $x^*=\lambda^*\delta$ 
to have a simpler notation, we have
\begin{eqnarray}
M(x)&=& -\frac{1}{2\delta^2}\sum_{n \geq 1} \frac{x^n}{n!}
\sum_{\ell \neq n}\frac{x^{*\ell}}{\ell !}
+ \frac{1}{\delta^2}\sum_{\ell \geq 2} \frac{x^{*\ell}}{ \ell !}
 \sum_{1\leq j < \ell}  \frac{x ^j}{ j!} =\nonumber\\
&=&-\frac{1}{2\delta^2}\sum_{n \geq 1} \frac{x^n}{n!}
\sum_{\ell > n}\frac{x^{*\ell}}{\ell !}-\frac{1}{2\delta^2}\sum_{n \geq 2} \frac{x^n}{n!}
\sum_{1\leq\ell < n}\frac{x^{*\ell}}{\ell !} \nonumber\\
&&+ \frac{1}{2\delta^2}\sum_{\ell \geq 2} \frac{x^{*\ell}}{ \ell !}
 \sum_{1\leq j < \ell}  \frac{x ^j}{ j!} 
+ \frac{1}{2\delta^2}\sum_{\ell \geq 2} \frac{x^{*\ell}}{ \ell !}
 \sum_{1\leq j < \ell}  \frac{x ^j}{ j!} ~.\nonumber\\
\end{eqnarray}

One can easily verify that the first and the third term sum up to zero, 
therefore it is
\begin{equation}
M(x)=-\frac{1}{2\delta^2}\sum_{n \geq 2} \frac{x^n}{n!}
\sum_{1\leq\ell < n}\frac{x^{*\ell}}{\ell !}
+ \frac{1}{2\delta^2}\sum_{\ell \geq 2} \frac{x^{*\ell}}{ \ell !}
 \sum_{1\leq j < \ell}  \frac{x ^j}{ j!}~,
\end{equation}
and, noticing that the expression above is nothing but the sum 
of the quantity $-\frac{1}{2\delta^2}\sum_{n \geq 2} \frac{x^n}{n!}
\sum_{1\leq\ell < n}\frac{x^{*\ell}}{\ell !}$ with its complex conjugate, 
we have $M(x)\in\Im$: this implies that only the odd terms (looking at the 
exponents $n+\ell$ as powers of $t\delta$) survive in the sum of the two 
series above. Finally, going back at Eq.~\eqref{e.G_A}, we get the
pure imaginary quantity 
\begin{eqnarray}\label{e.Gdelta_A}
G_\delta(t)&=&i{\Big\lbrace}-\frac{1}{\delta^2}{\Big[} -2t\delta \cos t\delta + 
2\sin t\delta +\nonumber\\
&&+ \Im{\rm m} {\Big(} \sum_{n \geq 2} \frac{(-it\delta)^n}{n!} \sum_{1\leq\ell < n}\frac{(it\delta)^\ell}{\ell !} {\Big)} {\Big]} {\Big\rbrace}\;,\nonumber\\
\end{eqnarray}
where the real time $t$ is back. This is the back-action term Eq.~\eqref{e.back-action} 
defined and analyzed in Sec.~\ref{ss.large-S}.

%\bibliographystyle{plain}
%\bibliographystyle{ieeetr}
%\bibliography{myBiblio_ac}

\begin{thebibliography}{28}
\expandafter\ifx\csname natexlab\endcsname\relax\def\natexlab#1{#1}\fi
\expandafter\ifx\csname bibnamefont\endcsname\relax
  \def\bibnamefont#1{#1}\fi
\expandafter\ifx\csname bibfnamefont\endcsname\relax
  \def\bibfnamefont#1{#1}\fi
\expandafter\ifx\csname citenamefont\endcsname\relax
  \def\citenamefont#1{#1}\fi
\expandafter\ifx\csname url\endcsname\relax
  \def\url#1{\texttt{#1}}\fi
\expandafter\ifx\csname urlprefix\endcsname\relax\def\urlprefix{URL }\fi
\providecommand{\bibinfo}[2]{#2}
\providecommand{\eprint}[2][]{\url{#2}}

\bibitem[{\citenamefont{Schlosshauer}(2007)}]{Schlosshauer07}
\bibinfo{author}{\bibfnamefont{M.}~\bibnamefont{Schlosshauer}},
  \emph{\bibinfo{title}{Decoherence and the Quantum-To-Classical Transition}},
  The Frontiers Collection (\bibinfo{publisher}{Springer},
  \bibinfo{year}{2007}).

\bibitem[{\citenamefont{Rivas and Huelga}(2012)}]{Rivas2012}
\bibinfo{author}{\bibfnamefont{A.}~\bibnamefont{Rivas}} \bibnamefont{and}
  \bibinfo{author}{\bibfnamefont{S.~F.} \bibnamefont{Huelga}},
  \emph{\bibinfo{title}{Open Quantum Systems: An Introduction}}
  (\bibinfo{publisher}{Springer Berlin Heidelberg}, \bibinfo{year}{2012}).

\bibitem[{\citenamefont{Breuer and Petruccione}(2002)}]{BreuerP02}
\bibinfo{author}{\bibfnamefont{H.~P.} \bibnamefont{Breuer}} \bibnamefont{and}
  \bibinfo{author}{\bibfnamefont{F.}~\bibnamefont{Petruccione}},
  \emph{\bibinfo{title}{The theory of open quantum systems}}
  (\bibinfo{publisher}{Oxford University Press}, \bibinfo{address}{Great
  Clarendon Street}, \bibinfo{year}{2002}).

\bibitem[{\citenamefont{Paladino et~al.}(2002)\citenamefont{Paladino, Faoro,
  Falci, and Fazio}}]{PaladinoEtal02}
\bibinfo{author}{\bibfnamefont{E.}~\bibnamefont{Paladino}},
  \bibinfo{author}{\bibfnamefont{L.}~\bibnamefont{Faoro}},
  \bibinfo{author}{\bibfnamefont{G.}~\bibnamefont{Falci}}, \bibnamefont{and}
  \bibinfo{author}{\bibfnamefont{R.}~\bibnamefont{Fazio}},
  \bibinfo{journal}{Phys. Rev. Lett.} \textbf{\bibinfo{volume}{88}},
  \bibinfo{pages}{228304} (\bibinfo{year}{2002}),
  \urlprefix\url{http://link.aps.org/doi/10.1103/PhysRevLett.88.228304}.

\bibitem[{\citenamefont{Lo~Franco et~al.}(2013)\citenamefont{Lo~Franco,
  Bellomo, Maniscalco, and Compagno}}]{LoFrancoEtal13}
\bibinfo{author}{\bibfnamefont{R.}~\bibnamefont{Lo~Franco}},
  \bibinfo{author}{\bibfnamefont{B.}~\bibnamefont{Bellomo}},
  \bibinfo{author}{\bibfnamefont{S.}~\bibnamefont{Maniscalco}},
  \bibnamefont{and} \bibinfo{author}{\bibfnamefont{G.}~\bibnamefont{Compagno}},
  \bibinfo{journal}{Int. J. Mod. Phys. B} \textbf{\bibinfo{volume}{27}},
  \bibinfo{pages}{1345053} (\bibinfo{year}{2013}).

\bibitem[{\citenamefont{Xu et~al.}(2013)\citenamefont{Xu, Sun, Li, Xu, Guo,
  Andersson, Lo~Franco, and Compagno}}]{XuEtal13}
\bibinfo{author}{\bibfnamefont{J.-S.} \bibnamefont{Xu}},
  \bibinfo{author}{\bibfnamefont{K.}~\bibnamefont{Sun}},
  \bibinfo{author}{\bibfnamefont{C.-F.} \bibnamefont{Li}},
  \bibinfo{author}{\bibfnamefont{X.-Y.} \bibnamefont{Xu}},
  \bibinfo{author}{\bibfnamefont{G.-C.} \bibnamefont{Guo}},
  \bibinfo{author}{\bibfnamefont{E.}~\bibnamefont{Andersson}},
  \bibinfo{author}{\bibfnamefont{R.}~\bibnamefont{Lo~Franco}},
  \bibnamefont{and} \bibinfo{author}{\bibfnamefont{G.}~\bibnamefont{Compagno}},
  \bibinfo{journal}{Nat Commun} \textbf{\bibinfo{volume}{4}},
  (\bibinfo{year}{2013}), \urlprefix\url{http://dx.doi.org/10.1038/ncomms3851}.

\bibitem[{\citenamefont{Matsumoto}(2015)}]{Matsumoto15}
\bibinfo{author}{\bibfnamefont{N.}~\bibnamefont{Matsumoto}},
  \emph{\bibinfo{title}{Classical Pendulum Feels Quantum Back-Action}},
  Springer Theses (\bibinfo{publisher}{Springer}, \bibinfo{year}{2015}), ISBN
  \bibinfo{isbn}{4431558802,9784431558804}.

\bibitem[{\citenamefont{Kippenberg and Vahala}(2008)}]{KippenbergV08}
\bibinfo{author}{\bibfnamefont{T.~J.} \bibnamefont{Kippenberg}}
  \bibnamefont{and} \bibinfo{author}{\bibfnamefont{K.~J.}
  \bibnamefont{Vahala}}, \bibinfo{journal}{Science}
  \textbf{\bibinfo{volume}{321}}, \bibinfo{pages}{1172} (\bibinfo{year}{2008}),
  ISSN \bibinfo{issn}{0036-8075},
  \urlprefix\url{http://science.sciencemag.org/content/321/5893/1172}.

\bibitem[{\citenamefont{Verhagen et~al.}(2012)\citenamefont{Verhagen,
  Deleglise, Weis, Schliesser, and Kippenberg}}]{VerhagenDWSK12}
\bibinfo{author}{\bibfnamefont{E.}~\bibnamefont{Verhagen}},
  \bibinfo{author}{\bibfnamefont{S.}~\bibnamefont{Deleglise}},
  \bibinfo{author}{\bibfnamefont{S.}~\bibnamefont{Weis}},
  \bibinfo{author}{\bibfnamefont{A.}~\bibnamefont{Schliesser}},
  \bibnamefont{and} \bibinfo{author}{\bibfnamefont{T.~J.}
  \bibnamefont{Kippenberg}}, \bibinfo{journal}{Nature}
  \textbf{\bibinfo{volume}{482}}, \bibinfo{pages}{63} (\bibinfo{year}{2012}),
  ISSN \bibinfo{issn}{0028-0836},
  \urlprefix\url{http://dx.doi.org/10.1038/nature10787}.

\bibitem[{\citenamefont{Lieb}(1973)}]{Lieb73}
\bibinfo{author}{\bibfnamefont{E.~H.} \bibnamefont{Lieb}},
  \bibinfo{journal}{Communications in Mathematical Physics}
  \textbf{\bibinfo{volume}{31}}, \bibinfo{pages}{327} (\bibinfo{year}{1973}),
  \urlprefix\url{http://projecteuclid.org/euclid.cmp/1103859040}.

\bibitem[{\citenamefont{Palma et~al.}(1996)\citenamefont{Palma, Suominen, and
  Ekert}}]{PalmaSE96}
\bibinfo{author}{\bibfnamefont{G.~M.} \bibnamefont{Palma}},
  \bibinfo{author}{\bibfnamefont{K.-A.} \bibnamefont{Suominen}},
  \bibnamefont{and} \bibinfo{author}{\bibfnamefont{A.~K.} \bibnamefont{Ekert}},
  \bibinfo{journal}{Proceedings of the Royal Society of London. Series A:
  Mathematical, Physical and Engineering Sciences}
  \textbf{\bibinfo{volume}{452}}, \bibinfo{pages}{567} (\bibinfo{year}{1996}),
  \urlprefix\url{http://rspa.royalsocietypublishing.org/content/452/1946/567.abstract}.

\bibitem[{\citenamefont{Cucchietti et~al.}(2010)\citenamefont{Cucchietti,
  Zhang, Lombardo, Villar, and Laflamme}}]{CucchiettiEtal10}
\bibinfo{author}{\bibfnamefont{F.~M.} \bibnamefont{Cucchietti}},
  \bibinfo{author}{\bibfnamefont{J.-F.} \bibnamefont{Zhang}},
  \bibinfo{author}{\bibfnamefont{F.~C.} \bibnamefont{Lombardo}},
  \bibinfo{author}{\bibfnamefont{P.~I.} \bibnamefont{Villar}},
  \bibnamefont{and} \bibinfo{author}{\bibfnamefont{R.}~\bibnamefont{Laflamme}},
  \bibinfo{journal}{Phys. Rev. Lett.} \textbf{\bibinfo{volume}{105}},
  \bibinfo{pages}{240406} (\bibinfo{year}{2010}),
  \urlprefix\url{http://link.aps.org/doi/10.1103/PhysRevLett.105.240406}.

\bibitem[{\citenamefont{Fernando~Casas}(2012)}]{CasasMN12}
\bibinfo{author}{\bibfnamefont{M.~N.} \bibnamefont{Fernando~Casas},
  \bibfnamefont{Ander~Murua}}, \bibinfo{journal}{Computer Physics
  Communications} \textbf{\bibinfo{volume}{20}} (\bibinfo{year}{2012}).

\bibitem[{\citenamefont{Zassenhaus}(1939)}]{Zassenhaus39}
\bibinfo{author}{\bibfnamefont{H.}~\bibnamefont{Zassenhaus}}, in
  \emph{\bibinfo{booktitle}{Abhandlungen aus dem Mathematischen Seminar der
  Universit\"at Hamburg}} (\bibinfo{year}{1939}), vol.~\bibinfo{volume}{13},
  pp. \bibinfo{pages}{1--100}.

\bibitem[{\citenamefont{Tavis and Cummings}(1968)}]{TavisC68}
\bibinfo{author}{\bibfnamefont{M.}~\bibnamefont{Tavis}} \bibnamefont{and}
  \bibinfo{author}{\bibfnamefont{F.~W.} \bibnamefont{Cummings}},
  \bibinfo{journal}{Phys. Rev.} \textbf{\bibinfo{volume}{170}},
  \bibinfo{pages}{379} (\bibinfo{year}{1968}),
  \urlprefix\url{http://link.aps.org/doi/10.1103/PhysRev.170.379}.

\bibitem[{\citenamefont{Yaffe}(1982)}]{Yaffe82}
\bibinfo{author}{\bibfnamefont{L.~G.} \bibnamefont{Yaffe}},
  \bibinfo{journal}{Rev. Mod. Phys.} \textbf{\bibinfo{volume}{54}},
  \bibinfo{pages}{407} (\bibinfo{year}{1982}),
  \urlprefix\url{http://link.aps.org/doi/10.1103/RevModPhys.54.407}.

\bibitem[{\citenamefont{Garraway}(2011)}]{Garraway11}
\bibinfo{author}{\bibfnamefont{B.~M.} \bibnamefont{Garraway}},
  \bibinfo{journal}{Phil. Trans. R. Soc. A} \textbf{\bibinfo{volume}{20}}
  (\bibinfo{year}{2011}).

\bibitem[{\citenamefont{Bennett et~al.}(2013)\citenamefont{Bennett, Yao,
  Otterbach, Zoller, Rabl, and Lukin}}]{BennettEtal13}
\bibinfo{author}{\bibfnamefont{S.~D.} \bibnamefont{Bennett}},
  \bibinfo{author}{\bibfnamefont{N.~Y.} \bibnamefont{Yao}},
  \bibinfo{author}{\bibfnamefont{J.}~\bibnamefont{Otterbach}},
  \bibinfo{author}{\bibfnamefont{P.}~\bibnamefont{Zoller}},
  \bibinfo{author}{\bibfnamefont{P.}~\bibnamefont{Rabl}}, \bibnamefont{and}
  \bibinfo{author}{\bibfnamefont{M.~D.} \bibnamefont{Lukin}},
  \bibinfo{journal}{Phys. Rev. Lett.} \textbf{\bibinfo{volume}{110}},
  \bibinfo{pages}{156402} (\bibinfo{year}{2013}),
  \urlprefix\url{http://link.aps.org/doi/10.1103/PhysRevLett.110.156402}.

\bibitem[{\citenamefont{H\"ark\"onen et~al.}(2009)\citenamefont{H\"ark\"onen,
  Plastina, and Maniscalco}}]{HaerkonenPM09}
\bibinfo{author}{\bibfnamefont{K.}~\bibnamefont{H\"ark\"onen}},
  \bibinfo{author}{\bibfnamefont{F.}~\bibnamefont{Plastina}}, \bibnamefont{and}
  \bibinfo{author}{\bibfnamefont{S.}~\bibnamefont{Maniscalco}},
  \bibinfo{journal}{Phys. Rev. A} \textbf{\bibinfo{volume}{80}},
  \bibinfo{pages}{033841} (\bibinfo{year}{2009}),
  \urlprefix\url{http://link.aps.org/doi/10.1103/PhysRevA.80.033841}.

\bibitem[{\citenamefont{Feng et~al.}(2015)\citenamefont{Feng, Zhong, Liu, Yan,
  Yang, Twamley, and Wang}}]{FengEtal15}
\bibinfo{author}{\bibfnamefont{M.}~\bibnamefont{Feng}},
  \bibinfo{author}{\bibfnamefont{Y.}~\bibnamefont{Zhong}},
  \bibinfo{author}{\bibfnamefont{T.}~\bibnamefont{Liu}},
  \bibinfo{author}{\bibfnamefont{L.}~\bibnamefont{Yan}},
  \bibinfo{author}{\bibfnamefont{W.}~\bibnamefont{Yang}},
  \bibinfo{author}{\bibfnamefont{J.}~\bibnamefont{Twamley}}, \bibnamefont{and}
  \bibinfo{author}{\bibfnamefont{H.}~\bibnamefont{Wang}}, \bibinfo{journal}{Nat
  Commun} \textbf{\bibinfo{volume}{6}},  (\bibinfo{year}{2015}),
  \urlprefix\url{http://dx.doi.org/10.1038/ncomms8111}.

\bibitem[{\citenamefont{Paladino et~al.}(2014)\citenamefont{Paladino, Galperin,
  Falci, and Altshuler}}]{PaladinoEtal14}
\bibinfo{author}{\bibfnamefont{E.}~\bibnamefont{Paladino}},
  \bibinfo{author}{\bibfnamefont{Y.~M.} \bibnamefont{Galperin}},
  \bibinfo{author}{\bibfnamefont{G.}~\bibnamefont{Falci}}, \bibnamefont{and}
  \bibinfo{author}{\bibfnamefont{B.~L.} \bibnamefont{Altshuler}},
  \bibinfo{journal}{Rev. Mod. Phys.} \textbf{\bibinfo{volume}{86}},
  \bibinfo{pages}{361} (\bibinfo{year}{2014}),
  \urlprefix\url{http://link.aps.org/doi/10.1103/RevModPhys.86.361}.

\bibitem[{\citenamefont{Benedetti et~al.}(2013)\citenamefont{Benedetti,
  Buscemi, Bordone, and Paris}}]{BenedettiEtal13}
\bibinfo{author}{\bibfnamefont{C.}~\bibnamefont{Benedetti}},
  \bibinfo{author}{\bibfnamefont{F.}~\bibnamefont{Buscemi}},
  \bibinfo{author}{\bibfnamefont{P.}~\bibnamefont{Bordone}}, \bibnamefont{and}
  \bibinfo{author}{\bibfnamefont{M.~G.~A.} \bibnamefont{Paris}},
  \bibinfo{journal}{Phys. Rev. A} \textbf{\bibinfo{volume}{87}},
  \bibinfo{pages}{052328} (\bibinfo{year}{2013}),
  \urlprefix\url{http://link.aps.org/doi/10.1103/PhysRevA.87.052328}.

\bibitem[{\citenamefont{Wold et~al.}(2012)\citenamefont{Wold, Brox, Galperin,
  and Bergli}}]{WoldEtal12}
\bibinfo{author}{\bibfnamefont{H.~J.} \bibnamefont{Wold}},
  \bibinfo{author}{\bibfnamefont{H.}~\bibnamefont{Brox}},
  \bibinfo{author}{\bibfnamefont{Y.~M.} \bibnamefont{Galperin}},
  \bibnamefont{and} \bibinfo{author}{\bibfnamefont{J.}~\bibnamefont{Bergli}},
  \bibinfo{journal}{Phys. Rev. B} \textbf{\bibinfo{volume}{86}},
  \bibinfo{pages}{205404} (\bibinfo{year}{2012}),
  \urlprefix\url{http://link.aps.org/doi/10.1103/PhysRevB.86.205404}.

\bibitem[{not()}]{notasulcampo}
\emph{\bibinfo{title}{It might seem that the same reasoning should also hold
  for the external field $h$, but that is actually a different issue: the role
  of $h$ is that of defining an energy scale for the magnetic system, and the
  free Hamiltonian stays physical in the $S\rightarrow\infty$ limit}}.

\bibitem[{\citenamefont{Mattis}(1981-)}]{MattisbookI}
\bibinfo{author}{\bibfnamefont{D.~C.} \bibnamefont{Mattis}},
  \emph{\bibinfo{title}{The theory of magnetism - Vol I}}
  (\bibinfo{publisher}{Springer-Verlag}, \bibinfo{address}{Berlin ; New York},
  \bibinfo{year}{1981-}).

\bibitem[{\citenamefont{Nielsen and Chuang}(2004)}]{Nielsen2004}
\bibinfo{author}{\bibfnamefont{M.~A.} \bibnamefont{Nielsen}} \bibnamefont{and}
  \bibinfo{author}{\bibfnamefont{I.~L.} \bibnamefont{Chuang}},
  \emph{\bibinfo{title}{{Quantum Computation and Quantum Information (Cambridge
  Series on Information and the Natural Sciences)}}}
  (\bibinfo{publisher}{Cambridge University Press}, \bibinfo{year}{2004}),
  \bibinfo{edition}{1st} ed., ISBN \bibinfo{isbn}{0521635039}.

\bibitem[{\citenamefont{Foti}(2015)}]{Foti_Master2015}
\bibinfo{author}{\bibfnamefont{C.}~\bibnamefont{Foti}}, Ph.D. thesis,
  \bibinfo{school}{University of Florence} (\bibinfo{year}{2015}).

\bibitem[{\citenamefont{Gilmore}(1972)}]{Gilmore72}
\bibinfo{author}{\bibfnamefont{R.}~\bibnamefont{Gilmore}},
  \bibinfo{journal}{Annals of Physics} \textbf{\bibinfo{volume}{74}},
  \bibinfo{pages}{391 } (\bibinfo{year}{1972}), ISSN \bibinfo{issn}{0003-4916},
  \urlprefix\url{http://www.sciencedirect.com/science/article/pii/0003491672901479}.

\end{thebibliography}

\end{document}